\documentclass[review]{elsarticle}

\usepackage[utf8]{inputenc}
\usepackage{lineno}
\usepackage{amsmath}
\usepackage{amsfonts}
\modulolinenumbers[1]

\usepackage[english]{babel}

\usepackage[letterpaper,top=2cm,bottom=2cm,left=3cm,right=3cm,marginparwidth=1.75cm]{geometry}

\usepackage{graphicx}
\usepackage[colorlinks=true, allcolors=blue]{hyperref}
\usepackage{bm}
\usepackage{algorithm}
\usepackage{algorithmic}
\usepackage{blindtext}
\usepackage{comment}
\usepackage{float}
\newcommand{\onum}[1]{\overline{{#1}}} 	
\newcommand{\unum}[1]{\underline{{#1}}}  	
\usepackage{color,soul}
\usepackage{setspace}
\journal{Journal}

\biboptions{authoryear}

\graphicspath{{./images/}}

\begin{document}

\begin{center}
{\Large Simulation Study of the Upper-limb Wrench Feasible Set with Glenohumeral Joint Constraints}\\ 
\end{center}

\textbf{Nasser REZZOUG, corresponding author}

Institut Pprime, CNRS - Université de Poitiers – ISAE-ENSMA - UPR 3346

11 Boulevard Marie et Pierre Curie

TSA 41123

86073 Poitiers CEDEX 9, France

Phone: 05 49 49 67 56

Email: nasser.rezzoug@univ-poitiers.fr\\

\textbf{Antun SKURIC}

Centre Inria de l'université de Bordeaux, Bordeaux, France\\

\textbf{Vincent PADOIS}

Centre Inria de l'université de Bordeaux, Bordeaux, France\\

\textbf{David DANEY}

Centre Inria de l'université de Bordeaux, Bordeaux, France\\


Type of submission: Full length article

Word count: 3482\\

\begin{frontmatter}






\newpage







\begin{abstract}
The aim of this work is to improve musculoskeletal-based models of the upper-limb Wrench Feasible Set i.e. the set of achievable maximal wrenches at the hand for applications in collaborative robotics and computer aided ergonomics. In particular, a recent method performing wrench capacity evaluation called the Iterative Convex Hull Method is upgraded in order to integrate non dislocation and compression limitation constraints at the glenohumeral joint not taken into account in the available models. Their effects on the amplitude of the force capacities at the hand, glenohumeral joint reaction forces and upper-limb muscles coordination in comparison to the original iterative convex hull method are investigated \textit{in silico}. The results highlight the glenohumeral potential dislocation for the majority of elements of the wrench feasible set with the original Iterative Convex Hull method and the fact that the modifications satisfy correctly stability constraints at the glenohumeral joint. Also, the induced muscles coordination pattern favors the action of stabilizing muscles, in particular the rotator-cuff muscles, and lowers that of known potential destabilizing ones according to the literature. 
\end{abstract}

\end{frontmatter}

 \newpage

\section{Introduction}
One fundamental assumption of human centered robot control is the ability to measure or estimate the physical capacity of humans to continuously adapts the robot assistance level based on the real-time need of its human counterpart ~\citep{Carmichael2013EstimatingPA,Figueredo2020,Pehlivan2016}. 
Moreover, within the framework of computer-aided design of workstations, the knowledge of human force capacities enables to evaluate to what extent a task is in adequacy with the capacities of the operators ~\citep{Figueredo2020,Perdeaux2010}, to define ergonomic criteria of discomfort~\citep{Maurice_2017} and to implement models of muscular fatigue~\citep{Ma2010,Savin2017,Savin2021}. Different methodologies have been proposed to assess human force capacities. Data based approaches rely on the availability of experimental data ~\citep{Khalaf2001,guenzkofer_2012_elbow,lannersten_1993_isometric,hall_2021_comparison,kotte_2018_normative,ladelfa2017} which are numerous but limited to particular postures.
Whitin the context of model based approaches, several algorithms using musculoskeletal models, predominantly of the upper-limb, are available ~\citep{skuric2022,Valero-Cuevas2009,Carmichael2013EstimatingPA,Ingram2016}. From posture, muscular forces and muscles moment arms, the set of achievable wrench at the hand called the Wrench Feasible Set (WFS) can be computed under the form of a convex polytope. Its main interest is that it allows an exhaustive evaluation of force capacities whatever the direction of force application or the posture.
The base assumption of all these algorithms is that muscular forces are the limiting factor. 
However, other factors can limit force capacities such as joint stability \citep{blache2017,blache2016,Ingram2016}, safety mechanisms implemented by the central nervous system for preventing joint injuries \citep{Assila_2021, Maurice_2017} or balance constraints \citep{Perdeaux2010}. When considering the upper-limb and the glenohumeral (GH) joint, the stability issue is of paramount importance ~\citep{lipitt1993,dickerson2008,Labriola2005,blache2017,Sinha_1999}. Indeed, this joint has the greatest range of motion of all the joints of the human body. This is due to the relative laxity of the shoulder ligaments, but especially to the shallow conformation of the humeral head on the glenoid fossa, which surface is relatively flat. The side effect is that the GH joint is prone to dislocation when the shear component of the GH Joint Reaction Force (JRF) 
exceeds a certain percentage of its compressive component~\citep{lipitt1993,dickerson2008}. To prevent this phenomenon, several muscles including the rotator cuff muscles act to stabilize the GH joint by orienting the JRF toward the middle of the glenoid fossa while pressing the humeral head against it~\citep{lipitt1993}. 
Therefore, the induced muscle coordination pattern may have a significant effect on strength capacities and, if not accounted for, may lead to a potential overestimation of the WFS. However, to our knowledge, the only works implementing GH non dislocation constraints considered the shoulder torque feasible set only~\citep{Ingram2016} or were used to asses shoulder muscles forces with inverse dynamics and static optimization~\citep{dickerson2008,blache2016,blache2017} or with EMG assisted optimization ~\citep{assila_2020,Assila_2021}.

To fill this gap and to improve models of WFS computed from musculoskeletal models, the recent iterative convex hull method (ICHM)~\citep{skuric2022,Skuric2023} is upgraded in order integrate two important limiting factors of force capacities  i.e. GH joint non dislocation and limitation of GH JRF compression component. The core of the ICHM consists in finding the vertices of the WFS by solving a linear programming problem. To ensure GH joint non dislocation, additional constraints take the form of inequalities involving the admissible ratio between shear and compressive components of the GH JRF~\citep{lipitt1993,dickerson2008,blache2016,Ingram2016}. Also additional constraints bound the GH JRF compression.
In this context, the objective of this work is to test \textit{in silico} the effects on the WFS of the ICHM implementation incorporating the GH constraints. In particular, the following points are verified:\
\begin{itemize}
    \item $P_1$: the decrease of the range of maximal achievable forces,
    \item $P_2$: the orientation of the GH JRF toward the middle of the glenoid fossa,
    \item $P_3$: the decrease of the amplitude of the GH JRF compression,
    \item $P_4$: the modifications of muscle force coordination pattern with an expected increase in the contribution of the rotator cuff~\citep{blache2017} and biceps muscles~\citep{Labriola2005} and a decrease in that of the deltoids and pectoralis major muscles~\citep{Labriola2005}.
\end{itemize}

They are tested by considering a validated musculoskeletal model of the upper-limb~\citep{Saul2015} in three different postures. 


\section{Methods}
\subsection{Wrench Feasible Set definition}\label{ch:wrench_feasabiliy_definition}
Let us consider a serial kinematic chain $\mathcal{C}$ with $n$ rotational degrees of freedom (dofs) representing the upper-limb. Its configuration is specified by the joint angle vector $\mathbf{q} \in \mathbb{R}^{n}$ and the joint velocity vector $\mathbf{\dot{q}} \in \mathbb{R}^{n}$.
$\mathbf{\tau} \in \mathbb{R}^{n}$ refers to the joint torque vector. The task space is of dimension $m=3$, the end effector position is defined by $\mathbf{x} \in \mathbb{R}^{m}$ and the end-effector wrench by $\bm{F} \in \mathbb{R}^{m}$. 
$\mathcal{C}$ is actuated by $d$ muscles ($d > n$) producing muscular tension force $\bm{t} \in \mathbb{R}^{d}$. The joint torque $\bm{\tau} \in \mathbb{R}^{n}$, generated by the muscle force $\bm{t} \in \mathbb{R}^d$ at joint configuration $\bm{q} \in \mathbf{R}^n$, can be calculated as follows:   
\begin{equation}
    \bm{\tau} = -L^{T}(\bm{q})\bm{t} 
    \label{eq:muscle_torqe_gen}
\end{equation}
$L(\bm{q})^T \in \mathbb{R}^{n \times d}$ is the transpose of the moment arm matrix, which is defined as the muscle length Jacobian relating the space of joint and muscle length velocities  
\begin{equation}
    \dot{\bm{l}} = L(\bm{q}) \dot{\bm{q}},\quad L_{ij}=\dfrac{\partial l_i}{\partial q_j}
\end{equation}
where $\dot{\bm{l}} \in \mathbf{R}^d$ is the muscle lengthening velocity and $\dot{\bm{q}} \in \mathbf{R}^n$ the joint angular velocity. The negative sign in equation (\ref{eq:muscle_torqe_gen}) makes the force applied in the length shortening direction of the muscle positive.

The muscles have a limited force capacity and each component $t_i=1..d$ of $\mathbf{t}$ is to lie within an interval $\left[\unum{t}_i,\onum{t}_i\right]$. Also,  $0\leq \unum{t}_i < \onum{t}_i$ because a muscle can only pull and not push.
The tension $t_i$ developed by muscle $i$ is defined according to a Hill type model~\citep{Saul2015}. 



A general achievable set of muscle-tendon forces  $\bm{t}\!\in\!\left[\unum{\bm{t}},\,\onum{\bm{t}}\right] $ forms a $d$-dimensional hyper-rectangle 
with side lengths equal to the ranges of each of the $d$ muscle forces. Using the moment arm matrix $L(\bm{q})$ and equation (\ref{eq:muscle_torqe_gen}) this hyper-rectangle can be projected into the $n$-dimensional space of joint torques $\bm{\tau}$, forming the convex polytope of achievable joint torques $\mathcal{P}_\tau$
\begin{equation}
    \mathcal{P}_\tau = \big\{ \bm{\tau}\! \in\! \mathbb{R}^{n} ~ | ~ \bm{\tau} =\! -L^{T}(\bm{q})\bm{t}, ~ \bm{t}\! \in\! \left[\unum{\bm{t}},\,\onum{\bm{t}}\right] \big\}
    \label{eq:torque_poly}
\end{equation}

Furthermore, the dual relationship of the $m$-dimensional Cartesian wrenches $\bm{F}$ and the generalised joint torques $\bm{\tau}$ is given through the Jacobian transpose matrix $\bm{J}^{T}(\bm{q})\bm{F}\! = \!\bm{\tau}$ and defines the achievable Cartesian feasible wrench:
\begin{equation}
    \mathcal{P}_F = \big\{ \bm{F}\! \in\! \mathbb{R}^{m} ~ | ~ J^{T}(\bm{q})\bm{F}\! =\! \bm{\tau}, ~ \bm{\tau}\! \in\! \mathcal{P}_\tau \big\}
    \label{eq:force_poly_1}
\end{equation}


Combining (\ref{eq:torque_poly}) and (\ref{eq:force_poly_1}) the implicit definition of $\mathcal{P}_F$ is
\begin{equation}
    \mathcal{P}_F = \left\{ \bm{F}\! \in\! \mathbb{R}^{m} ~ | ~ \bm{J}^{T}(\bm{q})\bm{F}\! =\! -\bm{L}^{T}(\bm{q})\bm{t}, \bm{t}\! \in\! \left[\unum{\bm{t}},\,\onum{\bm{t}}\right]\right\}
    \label{eq:full_polytope}
\end{equation}


\subsection{Outline of the ICHM algorithm}\label{subsec:lp_ICHM}



The ICHM is a recent algorithm for the feasibility set analysis of a generic class of linear algebra problems. It is suited for the determination of the feasible Cartesian WFS $\mathcal{P}_F$ associated to a musculoskeletal model of the human upper limb (\ref{eq:full_polytope})~\citep{skuric2022}. It is based on two processes. The first one consists in finding the vertices 
of $\mathcal{P}_F$ by solving a linear programming (LP) problem for a chosen set of directions $\bm{c} \in R^m$ in the Cartesian task space. The second one consists in computing the convex hull of the current vertices in order to define a new directions set $\bm{c}$ orthogonal to the current polytope faces. This process is iterated until the distance between the new vertices and the current polytope faces are below a predefined threshold. 
In the case of the WFS $\mathcal{P}_F$ determination, the LP problem solved during the vertex search for appropriately chosen $\bm{c}$ directions is the following:


\begin{equation}
\begin{aligned}
    \max_{\bm{t}} \quad &  -\bm{c}^T\bm{J}^{T+}(\bm{q})\bm{L}^T(\bm{q})\bm{t} \\
     \textrm{s.t.} \quad &  \bm{V}_2^T\bm{L}^T(\bm{q})\bm{t} = \bm{0} \\
          & \unum{\bm{t}} \leq \bm{t} \leq \onum{\bm{t}} \\
\end{aligned}
\label{eq:lin_prog}
\end{equation}

$\bm{V}_2 \in \mathbb{R}^{n \times (n-r)}$ forms an orthonormal basis of $\mathcal{K}er(\mathbf{J}(\bm{q})) = \mathcal{I}m(\mathbf{J}^T(\bm{q}))^\perp$. It is obtained from a singular value decomposition of $\bm{J}(\bm{q})=\bm{U}\bm{\Sigma}\bm{V}^T$ where $\bm{V} = [\bm{V_1}, \bm{V_2}]^T$. The null space basis of $\mathbf{J}(\bm{q})$ is spanned by the columns of $\bm{V}$ corresponding to singular values equal to zero and stored in $\bm{V}_2$. $r$ is the rank of $\mathbf{J}(\bm{q})$ and $\bm{J}^{T+}(\bm{q})$ is the Moore-Penrose pseudo-inverse of the $\bm{J}(\bm{q})^T$. 

The LP problem (\ref{eq:lin_prog}) finds a vertex or face of $\mathcal{P}_F$ furthest to a face of the current polytope which normal is $\bm{c}$ by finding the maximum of the projection of a wrench $\bm{F}=-\bm{J}^{T+}(\bm{q})\bm{L}^T(\bm{q})\bm{t}$ on the $\bm{c}$ direction. The equality constraints $ \bm{V}_2^T\bm{L}^T(\bm{q})\bm{t} = \bm{0}$ ensure that $\bm{L}^T(\bm{q})\bm{t} \in \mathcal{I}m(\bm{J}^T)$ and therefore that $\bm{F}=-\bm{J}^{T+}(\bm{q})\bm{L}^T(\bm{q})\bm{t}$ is an exact solution for the wrench $\bm{F}$ in the equation $\bm{J}^{T}(\bm{q})\bm{F}\! =\! -\bm{L}^{T}(\bm{q})\bm{t}$. In other words, the equality constraints guarantee that the wrench $\bm{F}$ can be generated by the joint torque $\bm{\tau}=-\bm{L}^T(\bm{q})\bm{t}$ such that $\bm{F}=\bm{J}^{T+}(\bm{q})\bm{\tau}$. The LP formulation is very flexible and allows to add new constraints relative to the GH non dislocation involving the JRF. 

\subsection{Adding GH joint non dislocation constraints to the vertex search algorithm of the ICHM}

\subsubsection{Formulation in terms of GH Joint Reaction Force}
Due to the particular conformation of the humerus and the glenoid fossa of the scapula, GH joint dislocation can occur if the JRF exerted by the humerus head on the the glenoid fossa is outside a particular irregular conic region (Figure \ref{fig:gh_dislocation_constraints}). Its boundaries can be defined by considering the admissible ratio $\mu_j$ between the shear and compression components of the JRF which are parallel and orthogonal to the plane of the glenoid fossa~\citep{dickerson2008,lipitt1993}, respectively. The interior of the polyhedral conic region is defined by a serie of affine constraints represented by half planes (Figure \ref{fig:gh_inequalities}). For a direction $j$ defined by an angle $\theta_j$ within the glenoid fossa plane $(\bm{y}_{GH}, \bm{z}_{GH})$, the constraint is expressed as following:


\begin{equation}
    \begin{bmatrix}
    \mu_j & cos\theta_j & sin\theta_j
    \end{bmatrix}\!^{GF}\bm{F}_{H/S} \leq 0
    \label{eq:disloc_j}
\end{equation}
where $^{GF}\bm{F}_{H/S}$ is the Joint Reaction Force exerted by the humerus head (H) on the scapula glenoid fossa (S) expressed in the frame of the glenoid fossa (superscript GF). The first component of $^{GF}\bm{F}_{H/S} = [F_n, F_{t1}, F_{t2}]$ is compression and the other two are shear forces. $F_n$ along $\mathbf{x}_{GF}$ is always negative due to the choice of the glenoid fossa frame (Figure \ref{fig:gh_dislocation_constraints}).

For a set of directions, a linearized version of the cone boundaries is obtained by stacking the constraints (\ref{eq:disloc_j}) in the matrix $\bm{C}_{DIS}$:

\begin{equation}
\label{eq:constr_disloc_jrf2}
\bm{C}_{DIS}
\!~^{GF}\bm{F}_{H/S} \leq 
\bm{0}
\end{equation}

----- please insert Figure \ref{fig:gh_dislocation_constraints} around here ----

----- please insert Figure \ref{fig:gh_inequalities} around here ----

\subsubsection{Incorporating muscular and external forces}


Let consider the static equilibrium of the humerus expressed in the global frame (denoted by the superscript 0). The weight of the segments is neglected.


\begin{equation}
\label{eq:hsteq}
^0\bm{F}_{S/H} + ^0\bm{F}_{F/H}+
\Sigma ^0\bm{F}_{M/H} = 0    
\end{equation}

$^0\bm{F}_{S/H}$ and $^0\bm{F}_{F/H}$ represents the forces exerted on the humerus by the scapula and forearm, respectively, $\Sigma ^0\bm{F}_{M/H}$ represents the action of forces (expressed in the global frame) of muscles acting on the humerus and crossing the GH joint. An external force $^0\bm{F}_{EXT}$ exerted at the hand is transmitted to the humerus according to the third Newton's law. Then, $^0\bm{F}_{F/H} =\! ^0\bm{F}_{EXT}$. $^0\bm{F}_{H/S}$ the Joint Reaction Force exerted by the humerus on the scapula in the global frame can be expressed as following:
\begin{equation}
    ^0\bm{F}_{H/S} = \Sigma ^0\bm{F}_{M/H} + \!^0\bm{F}_{EXT}
    \label{eq:statequil}
\end{equation}


The contribution of muscular forces to the GH JRF, $\Sigma ^0\bm{F}_{M/H}$, can be replaced by $^0\bm{C}_{JRF}(\bm{q})\bm{t}$ with $^0\bm{C}_{JRF}\in \mathbb{R}^{m \times d}$. Each column of $^0\bm{C}_{JRF}(\bm{q})$, called the force direction matrix~\citep{Ingram2016}, stores the contribution $\bm{n}_i$ of a 1N force of muscle $i$ to the GH JRF:
\begin{equation}
^0\bm{C}_{JRF}(\bm{q}) = 
\begin{bmatrix}
    \bm{n}_1 & ... & \bm{n_i} & ... & \bm{n_d}
\end{bmatrix}
\quad i = 1..d 
\end{equation}

Combining the action of muscles and that of $\!^0\bm{F}_{EXT}$, the GH JRF expression in the global frame is : 

\begin{equation}
\label{eq:c_jrf}
^0\bm{F}_{H/S} = ^0\bm{C}_{JRF}(\bm{q})\bm{t}+\!^0\bm{F}_{EXT}    
\end{equation}



It is considered that  $\!^0\bm{F}_{EXT}$ acting at the hand is counterbalanced by one element $\bm{F} \in \mathcal{P}_F$. Since the equality constraints in the LP (\ref{eq:lin_prog}) ensure that $\bm{L}^T(\bm{q})\bm{t} \in \mathcal{I}m(\bm{J}^T)$; for a certain $\bm{t}$, $\!^0\bm{F}_{EXT}$ can be expressed as:
\begin{equation}
\label{eq:f_jplust_N_F}
    \!^0\bm{F}_{EXT} = \bm{J}^{T+}(\bm{q})\bm{L}^T(\bm{q})\bm{t}
\end{equation}

Then (\ref{eq:c_jrf}) becomes:
\begin{equation}
\label{eq:c_jrf3}
^0\bm{F}_{H/S} = (^0\bm{C}_{JRF}(\bm{q})+\bm{J}^{T+}(\bm{q})\bm{L}^T(\bm{q}))\bm{t}    
\end{equation}


Expressed in the frame of the glenoid fossa (\ref{eq:c_jrf3}) becomes:
\begin{align}
\label{eq:c_jrf2}
\begin{split}
    ^{GF}\bm{F}_{H/S} =\! ^{GF}\bm{R}_0\!~^{0}\bm{F}_{H/S}\\
    ^{GF}\bm{F}_{H/S} =\! ^{GF}\bm{R}_0(^0\bm{C}_{JRF}(\bm{q})+\bm{J}^{T+}(\bm{q})\bm{L}^T(\bm{q}))\bm{t}
\end{split}
\end{align}

where $^{GF}\bm{R}_0$ is the rotation matrix from the glenoid fossa frame to the global frame.

\subsubsection{Constraints and LP formulation}

The GH non dislocation constraints are obtained by substituting $\!~^{GF}\bm{F}_{H/S}$ in (\ref{eq:constr_disloc_jrf2}) by its expression in  (\ref{eq:c_jrf2}):

\begin{align}
\label{eq:dis_constraints}
\begin{split}
\bm{C}_{DIS}~\!^{GF}\bm{R}_0(^0\bm{C}_{JRF}(\bm{q})+\bm{J}^{T+}(\bm{q})\bm{L}^T(\bm{q}))\bm{t} \leq \bm{0}\\
\bm{C}\bm{t} \leq \bm{0}
\end{split}
\end{align}

In addition, the amplitude of the first component of $^{GF}\bm{F}_{H/S}$ corresponding to compression is kept under a maximum value by considering the following constraint: 

\begin{equation}
\label{eq:limit_comp}
    \bm{D}\bm{t}-f_{max} \leq 0
\end{equation}

where $\bm{D}$ is the first row of $^{GF}\bm{R}_0(^0\bm{C}_{JRF}(\bm{q})+\bm{J}^{T+}(\bm{q})\bm{L}^T(\bm{q}))$ and $f_{max}$ the maximum allowed compressive force.

Finally, the formulation of the LP problem incorporating GH non dislocation constraints and limitation of GH JRF compression is:

\begin{equation}
\begin{aligned}
    \max_{\bm{t}} \quad &  -\bm{c}^T\bm{J}^{T+}(\bm{q})\bm{L}^T(\bm{q})\bm{t} \\
     \textrm{s.t.} \quad &  \bm{V}_2^T\bm{L}^T(\bm{q})\bm{t} = \bm{0} \\
     \quad & \bm{C}\bm{t}\leq \bm{0} \\
           & \bm{D}\bm{t}-f_{max} \leq 0\\
           & \unum{\bm{t}} \leq \bm{t} \leq \onum{\bm{t}} \\
\end{aligned}
\label{eq:lin_prog_non_disloc_constr}
\end{equation}

\subsection{Implementation details}
The seven degrees of freedom and 50 muscles Opensim upper-limb model proposed in~\citep{Saul2015} was used. The Jacobian matrix, muscles moment arm matrix, maximal and minimal muscle forces were obtained with the Matlab Opensim API.

Using the \texttt{calcReactionOnParentExpressedInGround} method of the Opensim \texttt{joint} object, each column $\bm{n}_{i}$ of $\bm{C}_{JRF}(\bm{q})$ was obtained by dividing $~^0\bm{F}_{H/S_i} = t_i\bm{n}_i$, the individual contribution of muscle $i=1..d$ to the GH JRF, by the value of the muscle tension $t_i$:

\begin{equation}
\label{eq:jrf_fmuscle_one_muscle}
    \bm{n}_{i} = \frac{^0\bm{F}_{H/S_i}}{t_i}
\end{equation}

The contribution $\bm{n}_i$ to the GH JRF was computed for each muscle by setting its activation $a$ to a random value in the interval $]0,1]$, to ensure that $t_i > 0$, and by setting the force of the other muscles to 0 (both passive and active components). In order to quantify the static GH JRF, the rotational inertia was set to huge values to prevent joint rotational acceleration and acceleration of the humerus center of gravity allowing to obtain individual contribution of a single muscle to the GH JRF in a static condition.

Since (\ref{eq:constr_disloc_jrf2}) was expressed in the frame of the glenoid fossa, the rotation matrix from this frame to the global frame was needed. Three markers were attached to the scapula of the chosen Opensim model (Figure \ref{fig:scapula_frame}) and the glenoid fossa frame was defined according to the method proposed in~\citep{blache2016}:

----- please insert Figure \ref{fig:scapula_frame} around here ----

\begin{align}
\label{eq:gh_frame}
\begin{split}
\bm{x}_{GF} = \frac{(\bm{gf}_2-\bm{gf}_3) \times (\bm{gf}_1-\bm{gf}_3)}{\lVert (\bm{gf}_2-\bm{gf}_3) \times (\bm{gf}_1-\bm{gf}_3) \rVert}\\
\bm{y}_{GF} = \frac{\frac{\bm{gf}_1+\bm{gf}_2}{2} -\bm{gf}_3}{\lVert \frac{\bm{gf}_1+\bm{gf}_2}{2} -\bm{gf}_3\rVert}\\
\bm{z}_{GF} = \bm{x}_{GF} \times \bm{y}_{GF}\\
^0\bm{R}_{GF} = \begin{bmatrix}\bm{x}_{GF} & \bm{y}_{GF} & \bm{z}_{GF} \end{bmatrix}\\
^{GF}\bm{R}_0 = {^0\bm{R}_{GF}}^T
\end{split}
\end{align}

The $\bm{x}_{GF}$ axis corresponded to the direction of the compressive component of the GH JRF which was always negative because the humeral head was pressing against the glenoid fossa. The $\bm{y}_{GF}$ and $\bm{z}_{GF}$ axes corresponded to shear forces because it was assumed that the glenoid fossa was almost flat.

The $\mu_i$ ratios were defined such that the constraint boundary was an ellipse with a ratio of 0.5 in the cranial-caudal directions and 0.3 in the anterior-posterior directions \citep{dickerson2008, lipitt1993}. $f_{max}$ was set to 1500N according to \citep{assila_2020, blache2017}.

\section{Results}
The WFS was computed using the ICHM algorithm with and without the GH constraints.
Three postures with different levels of arm abduction and elbow flexion were considered (Table \ref{tab:mod_joint_angles} and Figure \ref{fig:ul_postures}). They were chosen such that a range of GH abduction was covered and also because the third posture corresponded to the so-called "apprehension test" for shoulder luxation testing~\citep{McMahon2002-am,McMahon2003-df}. The forearm pronation-supination, wrist flexion-extension and radial-ulnar deviation were set to 0°.\\

----- please insert Table \ref{tab:mod_joint_angles} around here ----\\

----- please insert Figure \ref{fig:ul_postures} around here ----\\

----- please insert Figure \ref{fig:polytope} around here ----\\

----- please insert Figure \ref{fig:comp_polytopes} around here ----\\

The results were divided in three parts. Firstly, the WFS with and without GH non dislocation constraints were compared to verify $P_1$.   
In order to assess the global difference between WFS with and without GH constraints, the following procedure was used. The two WFS were superimposed. A set of rays originating from the center of the WFS with GH constraints was constructed. They were defined by a set of direction $\bm{v}(r,\theta, \varphi)$, with r = 1, $\theta \in$ [0, 360°] and $\varphi \in$ [0, 180°], which spherical coordinates vary with an increment of 1°. Each ray intersects \citep{Tuszynski2018} with both WFS and the "length" (in N) of the segment between the two points of intersection was computed for each ray. The graphical representation consisted of a sphere projected in 2D whose surface was colored according to the "length" of the segments corresponding to the $\bm{v}$ direction (Figure \ref{fig:comp_polytopes}). In addition, the difference between the minimal and maximal forces for each polytope and the segments "length" for some particular directions in the global reference frame (front, back, up, down, right and left) was also assessed (Table \ref{tab:diff_force_constraints}). 

In the second part, the GH JRF corresponding to each vertex of the WFS preventing GH dislocation was obtained from (\ref{eq:c_jrf3}) and the constraints satisfaction was checked by computing the ratio between shear and compressive components ($P_2$). The limit of compression forces was also checked ($P_3$). 

In the third part, the coordination of the shoulder muscles was investigated to verify $P_4$. A particular attention was paid to the forces of the deltoid and pectoralis major muscles on one side and rotator cuff and biceps muscles on the other side, which were known to have a destabilizing and stabilizing action on the GH joint, respectively \citep{Labriola2005,Sinha_1999}.

\subsection{Effect of GH joint contraints on the WFS}
The GH constraints reduced the WFS amplitude. The maximal forces were reduced by 66, 45 and 71 N for Posture 1, 2, and 3 respectively. Also, the maximal reduction were noticed when both non dislocation and compression constraints were ON. For posture 1; the main reductions were for front, up and left (55, 28 and 20 N respectively). For posture 2 it was front, left and down (124, 99 and 45 N, respectively) and for posture 3 front and down (66 and 24N). The global difference, of the sphere projected in 2D is given in Figure \ref{fig:polytope}.\\

----- please insert Table \ref{tab:diff_force_constraints} around here ----\\


\subsection{GH Constraints satisfaction}
 The ratio of the GH JRF shear forces divided by the compressive component superimposed with the boundary of the GH non dislocation constraints for the three chosen postures are presented in Figure~\ref{fig:gh_constraints}. Without the non dislocation constraints, the developed muscular forces conducted to a dislocation of the GH joint for 61$\%$, 55$\%$, and 76$\%$ of the vertices of the corresponding WFS for postures 1, 2, and 3, respectively (ratios depicted by red circle lying outside of the GH boundaries). When GH constraints were ON, these percentages dropped to 5$\%$, 2$\%$, and 3$\%$. For posture 1, the GH JRF vectors are directed upward (Figure \ref{fig:gh_constraints}.A), anteriorly for postures 2 and in the anterior-inferior direction for posture 3 (Figures \ref{fig:gh_constraints}.B and \ref{fig:gh_constraints}.C). When the GH constraints are active, the ratios between shear and compressive lye inside the constraints boundaries validating $P_2$. The JRF depicted in Figure \ref{fig:jrf} confirms the orientation of the dislocating JRF and show the effect of the constraints limiting the amplitude of compressive forces (blue thick arrows in Figure \ref{fig:jrf}).\\



----- please insert Figure \ref{fig:gh_constraints} around here ----\\

----- please insert Figure \ref{fig:jrf} around here ----\\

----- please insert Figure \ref{fig:muscle_coord} around here ----\\

----- please insert Figure \ref{fig:muscle_coord_compression} around here ----\\

\subsection{Muscle force coordination} 
The effect of GH constraints on shoulder muscles force coordination was investigated. The pectoralis major, anterior, middle and posterior parts of deltoid were considered for their destabilizing effect and the rotator cuff muscles (subscapularis, intraspinatus, supraspinatus and teres minor) and biceps (for high GH abduction values) for their stabilizing contribution~\citep{Labriola2005,Sinha_1999}. Firstly, the maximum compression constraint was relaxed in order to compare the sole effect of the non dislocation constraint i.e. the direction of the GH JRF within the glenoid fossa. For the WFS without GH constraints, the mean and standard deviation of muscular forces was computed separately for the vertices conducting to GH dislocation (red circles in Figure \ref{fig:gh_constraints}) and those not conducting to GH dislocation (green stars in Figure \ref{fig:gh_constraints}). The mean and standard deviation of muscular forces was assessed for all the vertices of the WFS with GH non dislocation constraints. The results are depicted in Figure \ref{fig:muscle_coord} for the three postures. 

\section{Discussion}
\subsection{Force amplitudes}
The consideration of GH contraints for avoiding dislocation and too excessive joint load had an effect on the WFS at the hand. A decrease of the maximal forces was noticed particularly in the anterior and mediolateral left directions and depended on the posture.

The non dislocation constraints had a moderate effect on the maximal forces at the hand but resulted in a modification of muscle coordination with a modulation of tension that ensured GH non dislocation. This illustrate the muscular redundancy implemented in the musculoskeletal model allowing a redistribution of muscular forces conducting stabilize the GH joint without affecting significantly the maximal forces. On the other hand, the maximum compression constraint contributed more substantially to the lowering of forces at the hand by a global decrease of all muscle tensions (Figure \ref{fig:muscle_coord_compression}). 
\subsection{Muscle coordination}
Apart from generating joint movements the muscles action is also directed toward the stabilization of the joints. This is especially true for the GH joint. Our results are coherent with those of the literature with a decrease of the force amplitude of the deltoids and pectoralis major in the three postures (Figure \ref{fig:muscle_coord}) with a particular implication of pectoralis muscle exemplified by the effect of its spasms in GH dislocation \citep{Sinha_1999}. In our results, it is particularly noticeable for this muscle with a decrease of the mean tension of 216 N, 575N and 746N corresponding to 27$\%$, 54$\%$ and 77$\%$ when comparing the WFS with GH non dislocation constraints to the WFS vertices conducting to dislocation for posture 1, 2, and 3 respectively. Moreover, in posture 3, corresponding to the apprehension test, it can be seen that without the proposed constraints, the JRF is directed toward the anterior-inferior direction. This observation is consistent with the literature indicating that GH dislocation occurs in this direction when performing forceful apprehension test on cadaveric upper-limbs~\citep{McMahon2002-am,McMahon2003-df}. 



\subsection{Limitations}
The present model has several limitations. Firstly, only muscular contribution to joint non dislocation was considered but not the static restraints such as capsulolabrum, articular surface geometry and intracapsular pressure. However, it is hypothesized that the non dislocation ratios consider their contribution in some way ~\citep{lipitt1993,dickerson2008}. Secondly, the WFS is computed in an isometric condition neglecting the segments weight and movement dynamics. To relax these hypotheses the equation relating muscle tension $\bm{t}$ and $\bm{F}$ should be replaced by the following one:
\begin{equation}
    J^{T}(\bm{q})\bm{F}  = -L(\bm{q})\bm{t} - \bm{\tau}_{b}
    \label{eq:bias_torque}
\end{equation}\
where $\bm{\tau}_{b}$ is the bias joint torque produced by the gravity and the model dynamics.
Results on muscles tension are global and are averaged for all the WFS vertices and therefore their standard deviation is high. It would be valuable to differentiate the results by considering the Cartesian directions. However, the global presentation was chosen for better synthesizing the results. 
Another limitation is that the results heavily depend on the quality of the musculoskeletal model in terms of muscle force and moment arm matrix. The tested model was generic without any scaling and was used to provide a proof of concept for the proposed method. 
Finally, an experimental validation is mandatory and will be conducted on the basis of the experiment carried out in \citep{rezzoug_2021_upper}.


\section{Conclusion}
The aim of this work was to propose an extension of the ICHM to assess the upper-limb WFS taking into account the limiting factor of GH joint stability. The adaptations of the ICHM were described and implemented. The results confirmed the lowering of the WFS due to the association of GH constraints on the JRF. The non dislocation constraints had an effect on the coordination of muscular forces while compression constraints limited muscle tensions. The change in muscle tension were investigated and in global agreement with the literature, the activity of some muscular groups was lowered or increased. Taking into account biomechanical constraints such as those implemented on the GH joint allow to improve models of musculoskeletal capacities for better simulating human behavior in various application such as collaborative robotics and ergonomics. 

\section{Conflict of Interest Statement}
The authors declare that they have no known competing financial interests or personal relationships that
could have appeared to influence the work reported in this paper.

\section{Acknowledgements}
This work has been partially funded by the BPI France Lichie project.

\bibliography{mybibfile}

\newpage

\newpage
\section{List of Tables}

\begin{table}[H]
\centering
\caption{Glenohumeral and elbow joint angles in degrees for the selected postures. EL: Elevation angle, SE: shoulder elevation, SR: Shoulder rotation, and EF: Elbow flexion.}
\label{tab:mod_joint_angles}
\begin{tabular}{|c|c|c|c|c|} 
\hline Posture & EL (°) & SE (°) & SR (°) & EF (°) \\
\hline Posture 1 & 10 & 20 & 0 & 70 \\ 
\hline Posture 2 &  0 & 70 & -30 & 90 \\  
\hline Posture 3 &  0 & 90 & -90 & 90 \\
\hline
\end{tabular}
\end{table}

\begin{table}[h!]
\centering
\caption{Force difference (N) between the WFS without and with GH constraints according to posture, constraints and directions. disloc : dislocation constraints, comp : compression constraints, min and max: directions of the minimal and maximal force of the WFS without GH constraints, respectively, front, back, up, down, right and left correspond to directions in the global reference frame.}
\label{tab:diff_force_constraints}
\begin{tabular}{|c|c|c|c|c|c|c|c|c|c|c|} 
\hline Posture & disloc & comp & min & max & front & back & up & down & right & left\\
\hline 1 & ON & OFF & 0 & 0 & 0 & 0 & 0 & 0 & 0 & 4 \\
\hline 1 & ON & ON & 9 & 66 & 55 & 3 & 28 & 0 & 10 & 20 \\
\hline 2 & ON & OFF & 1 & 0 & 22 & 2 & 0 & 13 & 1 & 46 \\
\hline 2 & ON & ON & 2 & 45 & 124 & 8 & 1 & 45 & 2 & 99 \\
\hline 3 & ON & OFF & 0 & 7 & 48 & 0 & -2 & 0 & 3 & -1 \\
\hline 3 & ON & ON & 1 & 41 & 66 & 2 & 7 & 24 & 7 & 2 \\
\hline
\end{tabular}
\end{table}

\newpage
\section{List of figures}
\begin{figure}[h!]
    \centering
    \includegraphics[scale=0.5]{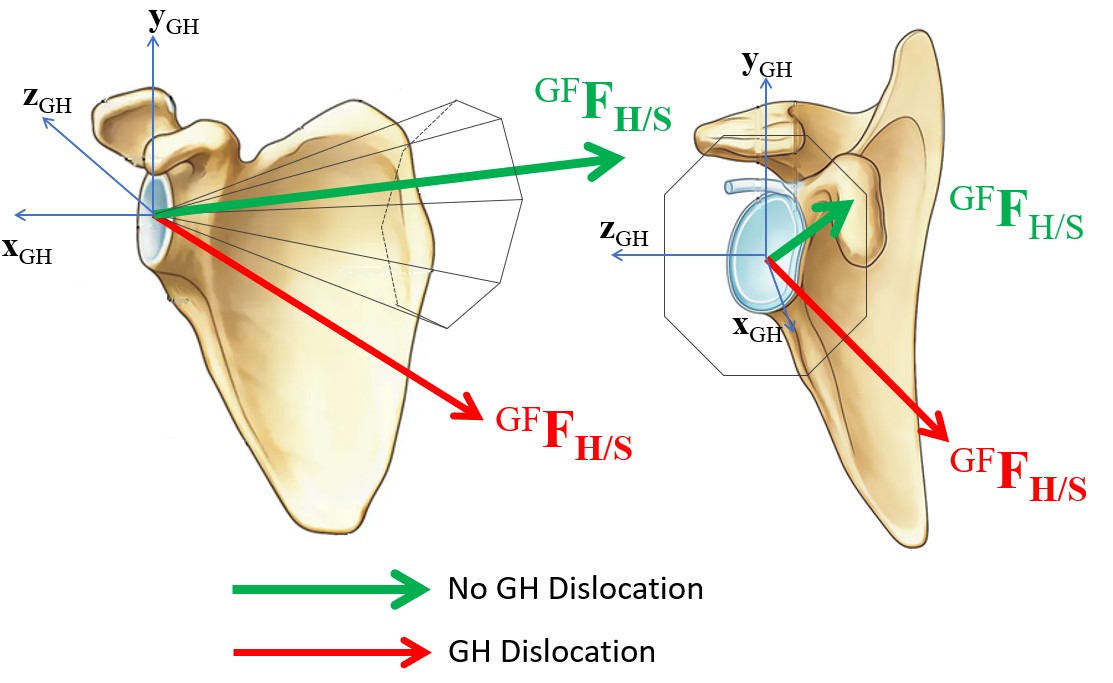}
    \caption{Glenohumeral dislocation constraints boundaries}
    \label{fig:gh_dislocation_constraints}
\end{figure}

\textbf{Figure \ref{fig:gh_dislocation_constraints}}: Glenohumeral dislocation constraints boundaries.\\

\begin{figure}[h!]
    \centering
    \includegraphics[scale=0.4]{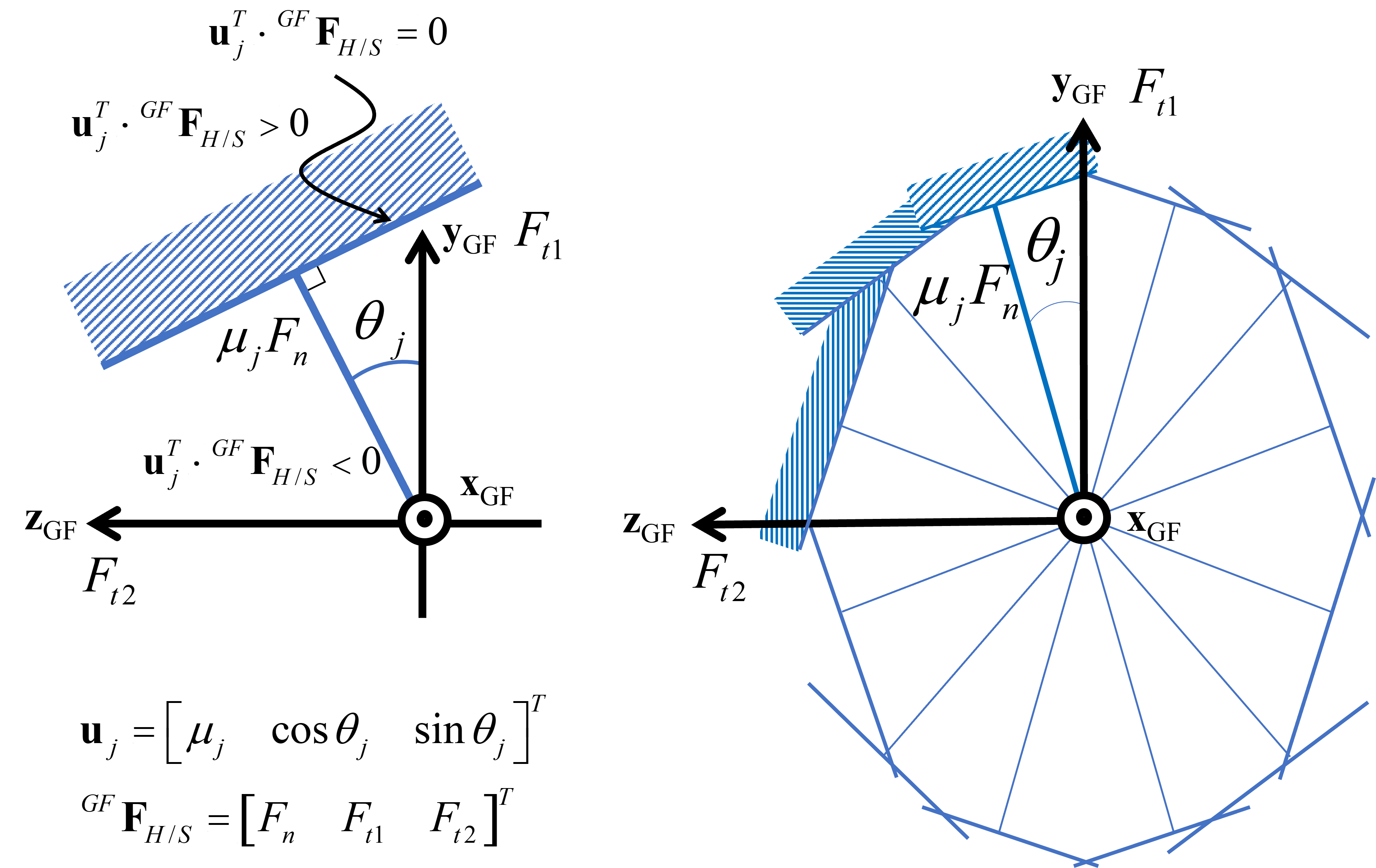}
    \caption{Glenohumeral dislocation constraints boundaries defined by a set of half-planes}
    \label{fig:gh_inequalities}
\end{figure}

\textbf{Figure \ref{fig:gh_inequalities}}: Glenohumeral dislocation constraints boundaries defined by a set of half-planes.\\

\begin{figure}[h!]
    \centering
    \includegraphics[scale=0.4]{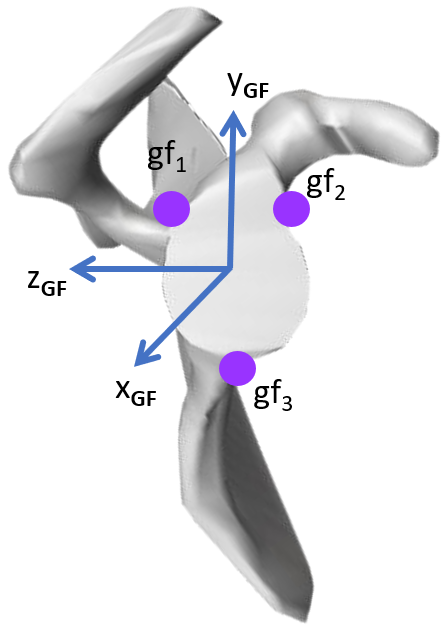}
    \caption{Glenoid fossa reference frame}
    \label{fig:scapula_frame}
\end{figure}

\textbf{Figure \ref{fig:scapula_frame}}: Glenoid fossa reference frame.\\

\begin{figure}[h!]
    \centering
    \includegraphics[scale=0.45]{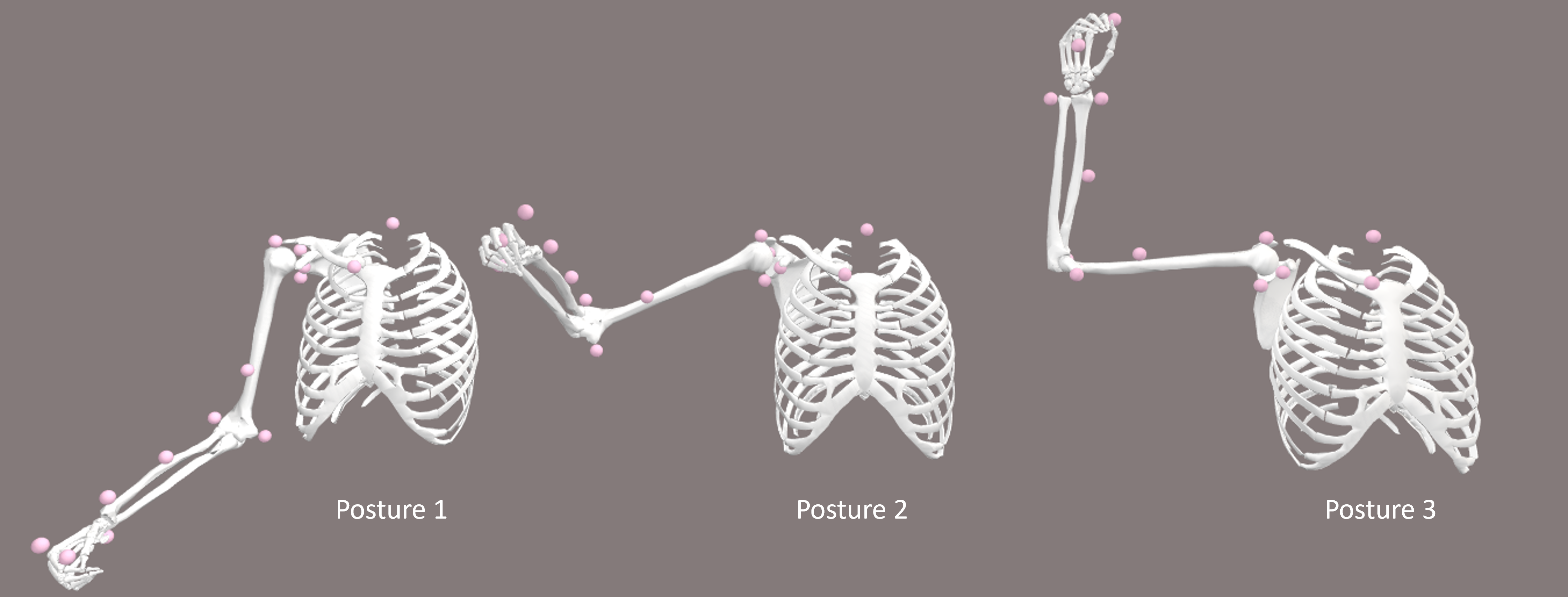}
    \caption{Considered upper-limb postures}
    \label{fig:ul_postures}
\end{figure}

\textbf{Figure \ref{fig:ul_postures}}: Considered upper-limb postures.\\

\begin{figure}[h!]
    \centering
    \includegraphics[scale=0.55]{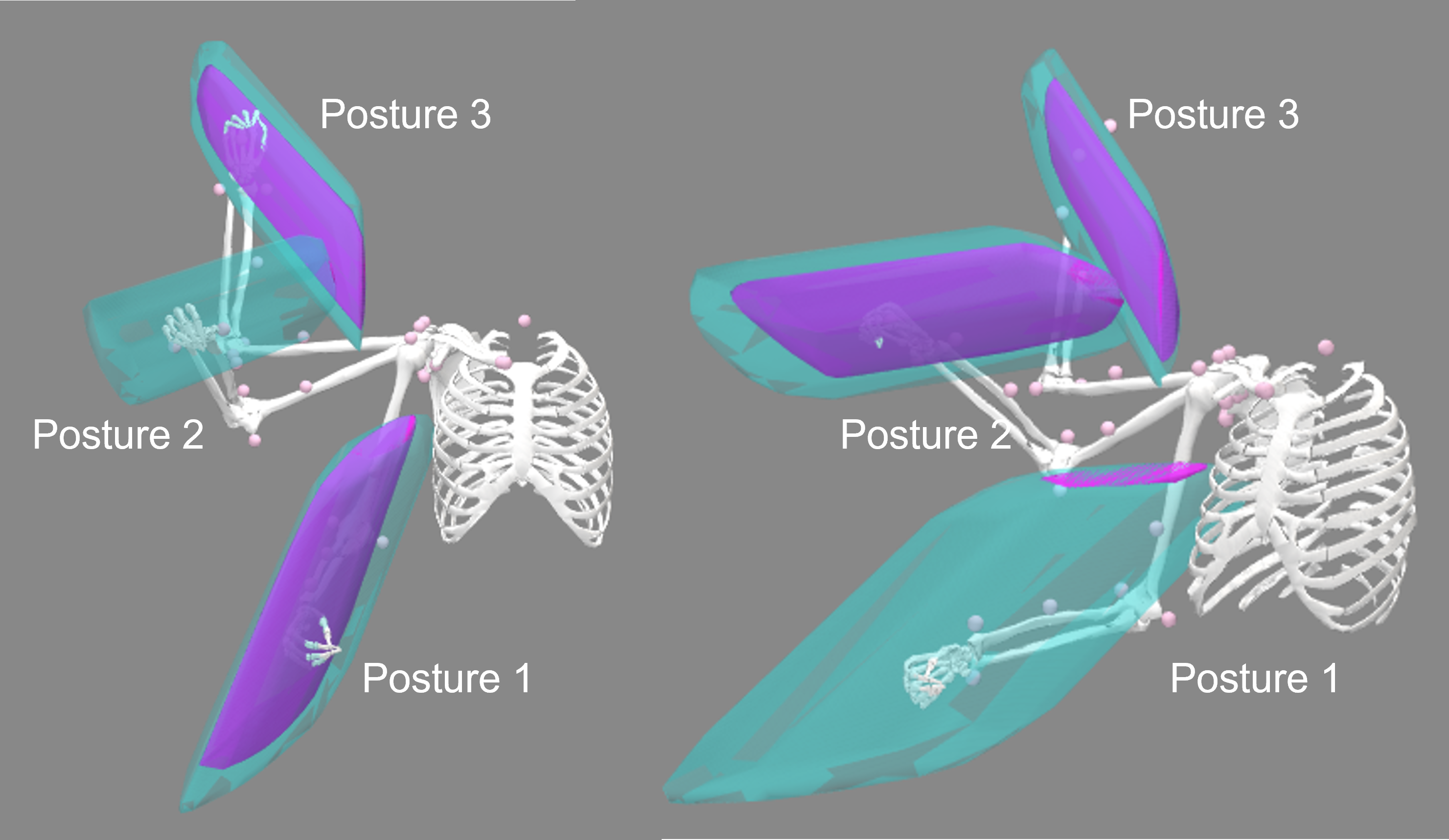}
    \caption{WFS and skeletal model representation for the three postures from two different points of view. Light (cyan): WFS without GH constraints, Dark (magenta): WFS with GH constraints.}
    \label{fig:polytope}
\end{figure}

\textbf{Figure \ref{fig:polytope}}: WFS and skeletal model representation for the three postures from two different points of view. Light (cyan): WFS without GH constraints, Dark (magenta): WFS with GH constraints.\\

\begin{figure}[h!]
    \centering
    \includegraphics[scale=0.48]{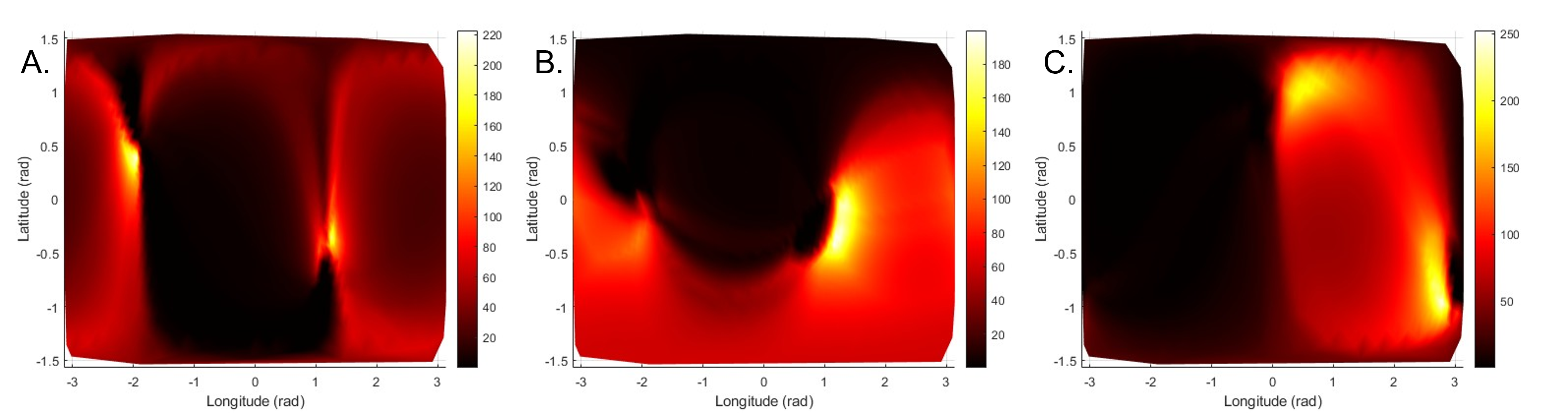}
    \caption{Difference (N) between polytopes according to Cartesian directions represented by their longitude and latitude (rad) for the three postures A.: posture 1, B. posture 2, and C. posture 3}
    \label{fig:comp_polytopes}
\end{figure}

\textbf{Figure \ref{fig:comp_polytopes}}: Difference (N) between polytopes according to Cartesian directions represented by their longitude and latitude (rad) for the three postures A.: posture 1, B. posture 2, and C. posture 3.\\

\begin{figure}[h!]
    \centering
    \includegraphics[scale=0.49]{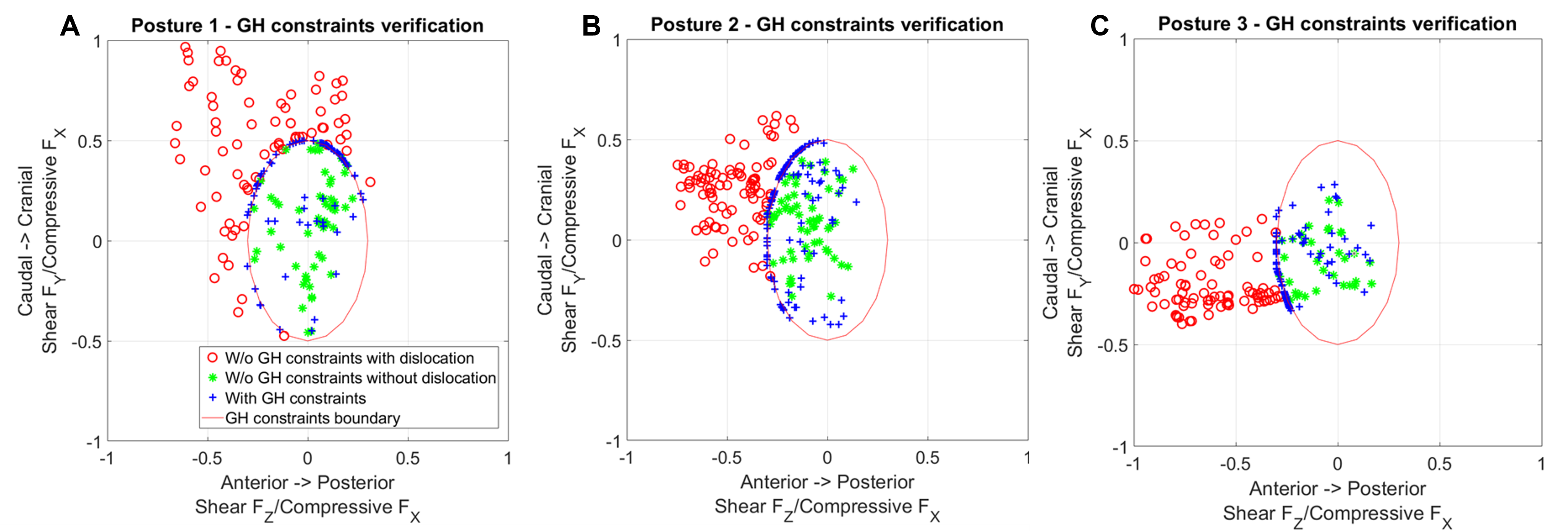}
    \caption{GH constraints verification: for one WFS vertex, the abscissa and ordinate of each 2D point correspond to the ratio of the shear component of the GH JRF around the Z and Y axis of the glenoid fossa frame, respectively, divided by the compressive component around the X axis. The red ellipse in each panel represent the GH non dislocation constraints boundaries. The red circles correspond to vertices of the WFS computed without GH constraints conducting to GH dislocation, the green stars to vertices of the WFS without GH constraints not conducting to GH dislocation, and finally the blue crosses correspond to vertices of a WFS with GH constraints on. Panels A, B, and C present the results for postures 1, 2, and 3 respectively.}
    \label{fig:gh_constraints}
\end{figure}

\textbf{Figure \ref{fig:gh_constraints}}: GH constraints verification: for one WFS vertex, the abscissa and ordinate of each 2D point correspond to the ratio of the shear component of the GH JRF around the Z and Y axis of the glenoid fossa frame, respectively, divided by the compressive component around the X axis. The red ellipse in each panel represent the GH non dislocation constraints boundaries. The red circles correspond to vertices of the WFS computed without GH constraints conducting to GH dislocation, the green stars to vertices of the WFS without GH constraints not conducting to GH dislocation, and finally the blue crosses correspond to vertices of a WFS with GH constraints on. Panels A, B, and C present the results for postures 1, 2, and 3 respectively\\

\begin{figure}[h!]
    \centering
    \includegraphics[scale=0.74]{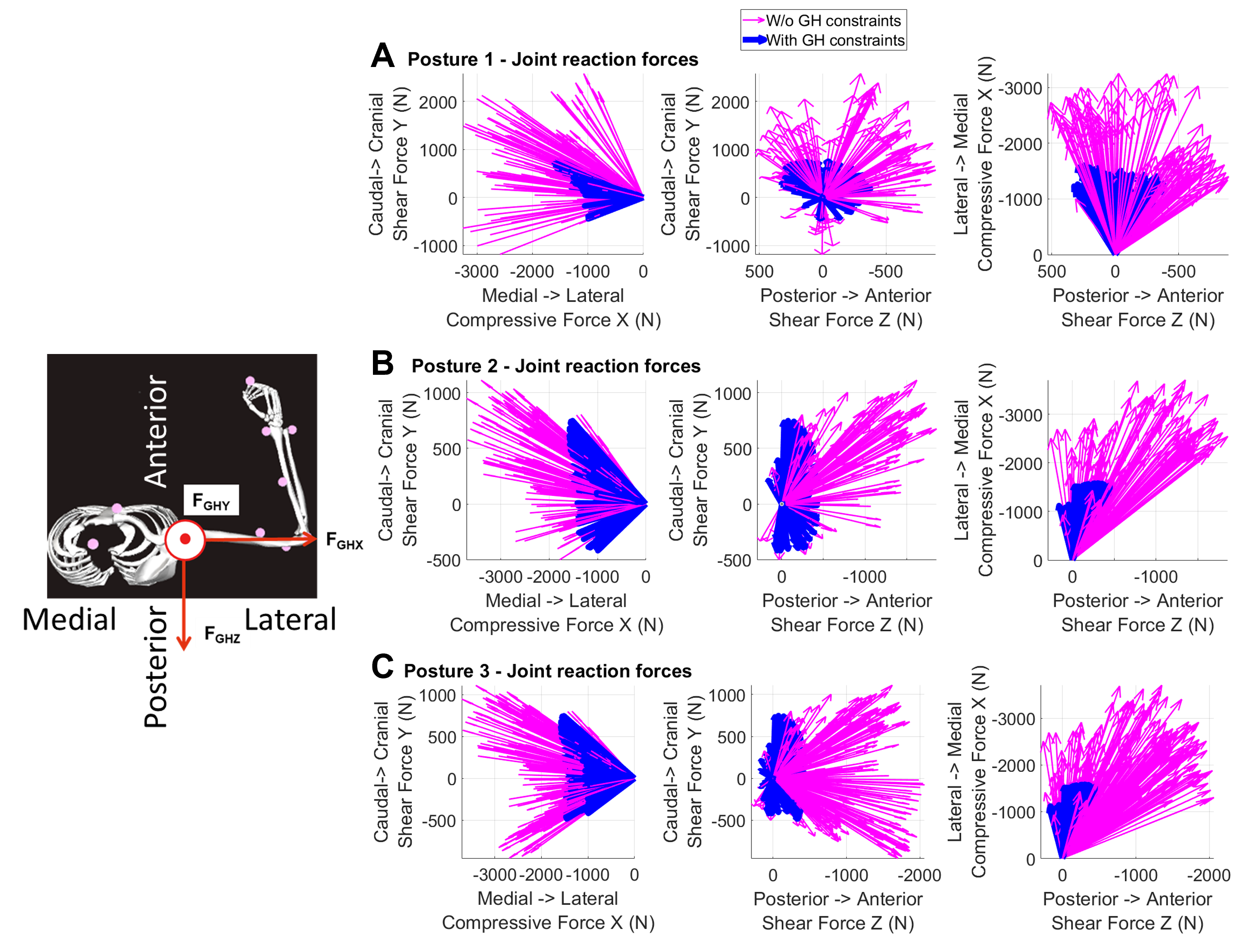}
    \caption{Each arrow depicts the GH JRF corresponding to one vertex of a WFS. The thin magenta arrows correspond to a WFS with GH constraints off and the thick blue arrows to a WFS with GH constraints on. Panels A, B, and C present the results for posture 1, 2, and 3 respectively. The left, middle and right figures present the results in the frontal, sagittal, and coronal planes,  respectively.}
    \label{fig:jrf}
\end{figure}

\textbf{Figure \ref{fig:jrf}}: Each arrow depicts the GH JRF corresponding to one vertex of a WFS. The thin magenta arrows correspond to a WFS with GH constraints off and the thick blue arrows to a WFS with GH constraints on. Panels A, B, and C present the results for posture 1, 2, and 3 respectively. The left, middle and right figures present the results in the frontal, sagittal, and coronal planes, respectively.\\

\begin{figure}[h!]
    \centering
    \includegraphics[scale=0.6]{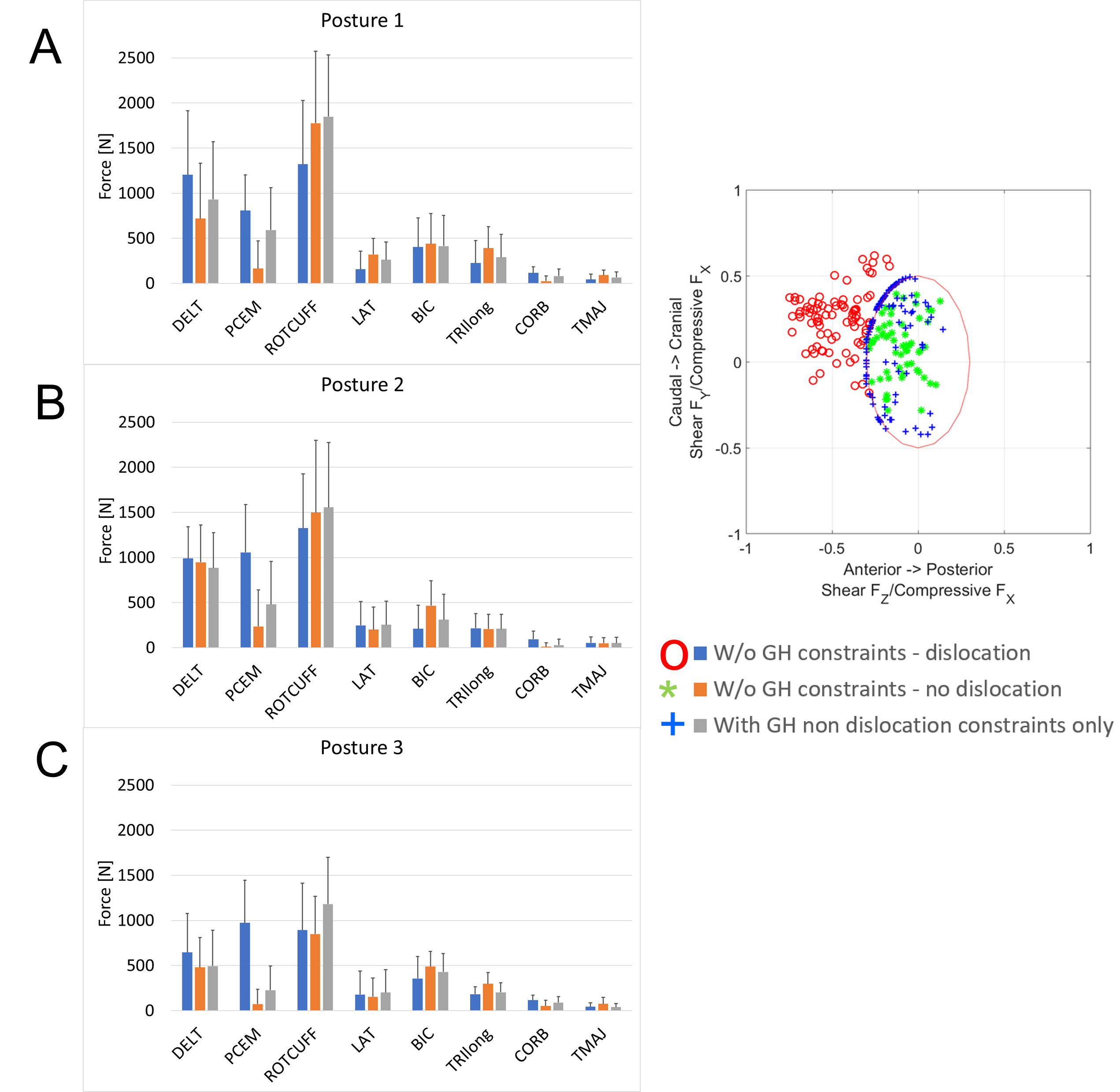}
    \caption{Mean and standard deviation of muscles tension according to posture and GH non dislocation constraints status. The GH maximum compression constraints are off. A, B, and C stand for posture 1, 2, and 3, respectively. PCEM: pectoralis major, DELT: sum of anterior, middle and posterior deltoid, ROTCUFF: sum of rotator cuff muscles: supraspinatus, infraspinatus, teres minor, and subscapularis, TMAJ: teres major, LAT: latissimus dorsi, CORB: coracobrachialis, TRIlong : long head of the triceps brachii, BIC: mean of biceps short and long heads.}  
    \label{fig:muscle_coord}
\end{figure}

\textbf{Figure \ref{fig:muscle_coord}}: Mean and standard deviation of muscles tension according to posture and GH non dislocation constraints status. The GH maximum compression constraints are off. A, B, and C stand for posture 1, 2, and 3, respectively. PCEM: pectoralis major, DELT: sum of anterior, middle and posterior deltoid, ROTCUFF: sum of rotator cuff muscles: supraspinatus, infraspinatus, teres minor, and subscapularis, TMAJ: teres major, LAT: latissimus dorsi, CORB: coracobrachialis, TRIlong : long head of the triceps brachii, BIC: mean of biceps short and long heads.\\

\begin{figure}[h!]
    \centering
    \includegraphics[scale=0.47]{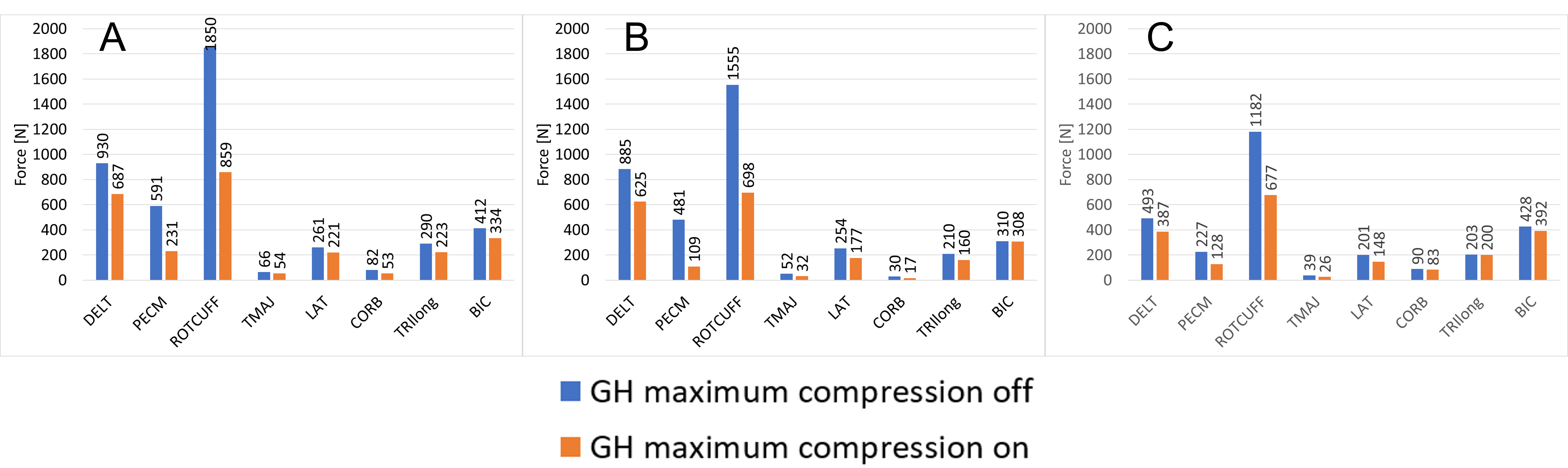}
    \caption{Muscles tension according to posture and GH compression constraints: blue (dark) constraint is off, orange (light) constraint is on. GH non dislocation constraints status is ON for both cases. A, B, and C stands for posture 1, 2, and 3, respectively. PCEM: pectoralis major, DELT: mean of anterior, middle and posterior deltoid, ROTCUFF: mean of rotator cuff muscles: supraspinatus, infraspinatus, teres minor, and subscapularis, TMAJ: teres major, LAT: latissimus dorsi, CORB: coracobrachialis, TRIlong : long head of the triceps brachii, BIC: mean of biceps short and long heads.}
    \label{fig:muscle_coord_compression}
\end{figure}

\textbf{Figure \ref{fig:muscle_coord_compression}}: Muscles tension according to posture and GH compression constraints: blue (dark) constraint is off, orange(light) constraint is on. GH non dislocation constraints status is ON for both cases. A, B, and C stands for posture 1, 2, and 3, respectively. PCEM: pectoralis major, DELT: mean of anterior, middle and posterior deltoid, ROTCUFF: mean of rotator cuff muscles: supraspinatus, infraspinatus, teres minor, and subscapularis, TMAJ: teres major, LAT: latissimus dorsi, CORB: coracobrachialis, TRIlong : long head of the triceps brachii, BIC: mean of biceps short and long
heads.\\

\newpage

\newpage

\newpage

\newpage

\newpage

\newpage

\newpage

\newpage

\newpage

\newpage

\end{document}